\documentclass[twocolumn,prl,aps,amsfonts]{revtex4}
\usepackage{graphicx}
\usepackage{psfrag, color}
\usepackage{verbatim}

\begin{document}

\title{Reentrant anisotropic phases in a two-dimensional hole system}

\author{M. J. Manfra$^1$, Z. Jiang$^2$, S. H. Simon$^1$, L. N. Pfeiffer$^1$, K. W. West$^1$, and A. M. Sergent$^1$}

\address{
 ${}^1$ Bell Laboratories, Lucent Technologies, Murray Hill, New Jersey
 \\ ${}^2$ NHMFL, Florida State University, Tallahassee, Florida}

\begin{abstract}
Anisotropic charge transport is observed in a two-dimensional
(2D) hole system in a perpendicular magnetic field at filling
factors $\nu$=7/2 and $\nu$=11/2 for temperatures below 150mK.  In stark contrast, the transport at $\nu$=9/2 is {\it isotropic} for
all temperatures. Our results for a two-dimensional hole system
differ substantially from 2D electron transport where no anisotropy
has been observed at $\nu$=7/2, the strongest anisotropy occurs at
$\nu$=9/2, and reentrant behavior is not evident.  We attribute
this difference to strong spin-orbit coupling in the hole
system.
\end{abstract}

\maketitle

For over a decade, investigation of the properties of half-filled
Landau levels of clean two-dimensional (2D) systems has been the
focus of intense research.  In two-dimensional {\it electron} systems
(2DESs) at half-filling, a surprisingly diverse set of ground
states has been uncovered.   In the N=0 Landau level (LL), at $\nu=1/2$
and $\nu=3/2$, compressible composite-fermion Fermi liquid states
are observed \cite{willett1,Olle}.   (Here $\nu=hcn/eB$ is the filling
factor with $B$ the magnetic field, and $n$ the carrier density).
In the N=1 Landau level, DC transport measurements have
convincingly demonstrated the presence of incompressible quantized
Hall states at $\nu$=5/2 and $\nu$=7/2 \cite{willett3,Pan1,Jim1}.
At half-filling in the $ 2 \leq $ N $\stackrel{<}{\sim} 5$ Landau
levels, electronic transport is anisotropic \cite{mike1,Ru1} and is
consistent with either a quantum smectic or nematic \cite{Kivelson}
(i.e., ``striped") phase.   While it is clear that all of these
phenomena derive from strong electron-electron interactions, the
exact relationship between the different ground states possible in
half-filled Landau levels and the sample parameters necessary to
stabilize one phase over another remain open questions \cite{Pan2,mike2,rezayi1}.  Access to a greater range of sample parameters than is currently available in high mobility 2DESs may enhance our understanding of these exotic states.

Transport studies of high mobility two dimensional {\it hole}
systems (2DHSs) offer a complimentary approach to the investigation
of correlation physics in 2D systems \cite{shayegan}.  The larger effective mass
of holes (m$_h$$\sim$0.5 vs. m$_e$$\sim$0.067, in units of the
free electron mass) reduces kinetic energy such that interactions play a more promiment role at a
given 2D density.  In addition, 2DHSs offer a ideal platform for the study of interesting spin
phenomena \cite{roland1} since spin-orbit coupling plays a significant role through mixing
of the light and heavy hole states.   Yet, for several practical
reasons, far fewer studies have been dedicated to the investigation of
half-filled Landau levels in 2DHSs \cite{mansour1}. First of all,
it is difficult to produce hole samples of sufficient quality such that correlations at half-filling are evident.  More importantly, the exploration of
anisotropic behavior in excited hole Landau levels has been
hindered by the presence of a significant mobility anisotropy at
zero magnetic field.   In the past, the highest mobility 2DHSs
have been grown on the (311)A orientation of GaAs where the zero
field mobility in the [$\bar{2}$33] direction often exceeds the
mobility along the [01$\bar{1}$] direction by a factor of 2 to 4
\cite{Heremans}. This zero field effect tends to obfuscate studies
of anisotropic behavior at higher magnetic fields.

In this Letter we present low temperature magnetotransport
measurements of a carbon-doped high mobility 2DHS \cite{manfra}
grown on the (100) surface of  GaAs where the zero field
mobility anisotropy is approximately 20\%.  (This residual anisotropy is also typical
in high mobility 2DESs).   At T=50mK we observe a pronounced
anisotropy in transport at filling factors $\nu$=7/2 and
$\nu$=11/2 while the transport at $\nu$=9/2 remains {\it
isotropic}.  At T=50mK, the resistance at $\nu$=7/2 in the [011]
direction exceeds the resistance in the [01$\bar{1}$] direction by
a factor \cite{Simon} of 16. The transport is extremely sensitive
to temperature. Isotropic transport is restored at $\nu$=11/2 for
temperatures greater than T$\sim$80mK and at $\nu=7/2$ for
temperatures greater than T$\sim$160mK. These results are the
first observation of anisotropic behavior in a 2D system in a
perpendicular magnetic field at $\nu$=7/2, a state charaterized by
a quantized Hall effect in clean 2D electron systems.  Our results
further indicate that 2D hole systems can support an unusual
alternating series of anisotropic/isotropic phases at half-filling
which is quite distinct from the known behavior of 2D electron
systems. We will interpret this new behavior as a result of strong spin-orbit coupling in the 2D hole system.

\begin{figure}
\includegraphics[width=\columnwidth]{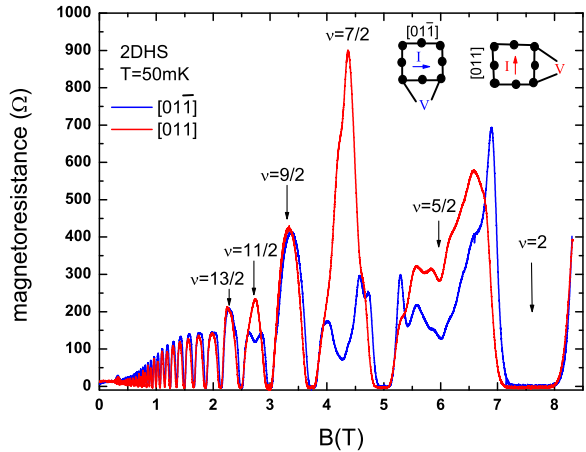}
 \caption{Overview
of magnetoresistance in our high-mobility two-dimensional hole
system at T=50mK.  The blue trace is measured with current flowing
in the [01$\overline{1}$] direction.  The red trace corresponds to
the magnetoresistance with the current flowing in the [011]
direction.  Anisotropic transport is clearly visible at $\nu$=11/2
and $\nu$=7/2 while $\nu$=9/2 remains isotropic.}
\end{figure}

The samples used in this experiment are 15nm-wide GaAs/AlGaAs
quantum wells grown on the (100) surface of GaAs by molecular beam
epitaxy.  The details of sample growth have been given
elsewhere \cite{manfra}.  The samples are symmetrically doped with
carbon, yielding a sheet density p=3.6$\times$10$^{11}$cm$^{-2}$
and mobility $\mu$=1$\times$10$^6$cm$^{2}$/Vs when cooled to
T=50mK in the dark.  Cyclotron resonance studies of similarly
grown carbon-doped quantum wells indicate that the hole effective
mass is large, $m_{h}\sim 0.5$ \cite{Han}.  The high mobility of
10$^{6}$cm$^{2}$/Vs in light of such a large effective mass
attest to the quality of our 2DHS.

\begin{figure}
\includegraphics[width=.9\columnwidth]{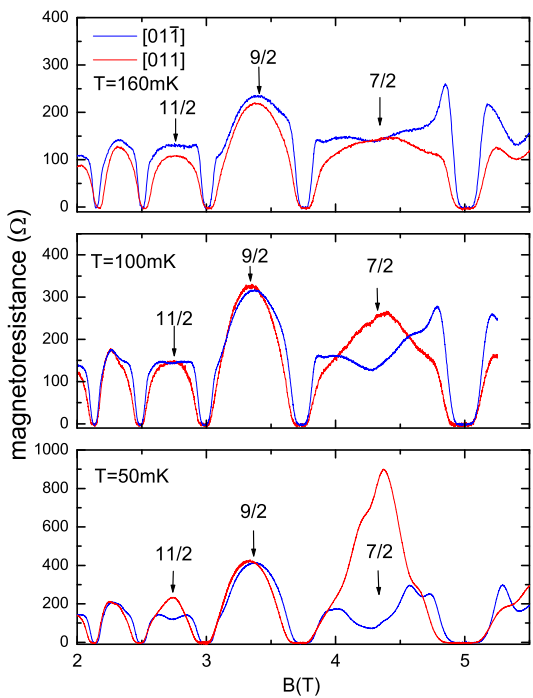}
 \caption{Magnetoresistance along [011] (red trace) and [01$\overline{1}$] (blue trace)
directions at temperatures T=160mK, T=100mK, and T=50mK.  The
resistance at $\nu$=9/2 increases slightly with decreasing
temperature but remains isotropic, while the anisotropies at
$\nu$=11/2  and $\nu$=7/2 increase in strength below T=100mK.}
\end{figure}

The samples are cleaved into 4mm by 4mm squares with eight In(Zn)
contacts placed symmetrically around the perimeter of the square.
The described transport experiments have been conducted with three
samples from two distinct wafers.  Similar anisotropic behavior is
observed in all samples.  Magnetotransport measurements are
performed in a dilution refrigerator with a base temperature
T=45mK using standard low frequency (11Hz) lock-in techniques. The
excitation current is kept to $\leq$ 10nA where no carrier heating
is observed at T=50mK.

Figure 1 presents an overview of transport in our 2DHS for filling
factor $\nu \geq$ 2 at T=50mK along the [011] and
[01$\overline{1}$] directions and contains the central findings of
this work.  At low magnetic field below B=2.5T the transport is
isotropic -- the zero magnetic field resistance ratio for this
sample is R$_{[011]}$/R$_{[01\overline{1}]}$ $\sim$ 1.4.  We note
that the resistances along [011] and [01$\overline{1}$] have not
been scaled to have equal amplitude in the low magnetic field regime.
At $\nu=11/2$ the first indication of anisotropic transport is
evident.  For current flow along the [011] direction a local
maximum in the longitudinal resistance is observed.  In the
[01$\overline{1}$] direction, a weak minimum is present.  This
anisotropy is extremely sensitive to temperature. While it is
clearly visible at T=50mK, the resistance at $\nu$=11/2 is
isotropic at T $\geq$ 80mK (see Fig. 2).  At $\nu$=9/2, the
resistance is {\it isotropic} at our base temperature.  At
$\nu$=7/2 the resistance again becomes anisotropic.  In the [011]
direction a strong peak in resistance is observed, while in the
[01$\overline{1}$] direction a strong minimum in resistance is
seen.  At T=50mK, the resistance ratio
R$_{[011]}$/R$_{[01\overline{1}]}$ reaches $\sim$ 16.

\begin{figure}
\includegraphics[width=.9\columnwidth]{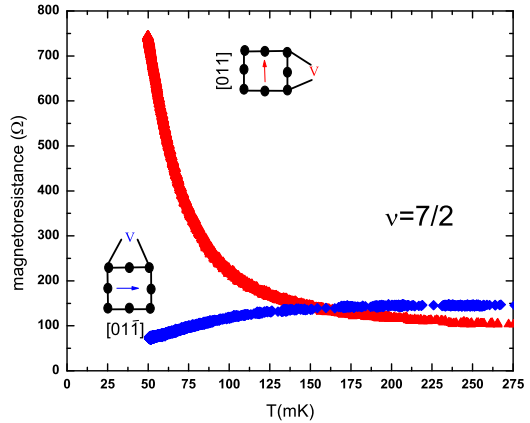}
 \caption{Temperature dependence of the resistance along the [011] direction (red) and
the [01$\overline{1}$] (blue) exactly at $\nu$=7/2.  Below
T$\sim$100mK, the resistance along [011] rises rapidly and is not
saturated at our base temperature T=50mK.  The resistance along
[01$\overline{1}$] continues to decrease down to T=50mK.}
\end{figure}

The temperature evolution of the magnetotransport for filling
factors between $\nu$=7 and $\nu$=3 is shown in Figure 2.  At
T=160mK, the resistance at $\nu$=11/2, 9/2, and 7/2 is largely
isotropic with little indication of the presence of correlation
effects.  At T=100mK, $\nu$=11/2 and $\nu$=9/2 remain isotropic,
while a developing anisotropy is visible at $\nu$=7/2.  The
resistance ratio R$_{[011]}$/R$_{[01\overline{1}]}$ at T=100mK at
$\nu$=7/2 is only $\sim$ 1.8.  Upon reducing the temperature to
T=50mK, the anisotropy at $\nu$=11/2 becomes clearly visible,
although its development is most probably limited by our
relatively high base temperature of 50mK \cite{mike1,mike2}.
Suprisingly, the resistance at $\nu$=9/2 remains isotropic.  At
$\nu$=7/2 no indication of a quantized Hall state is present,
R$_{xy}$ remains linear such that no developing plateau is evident
(not shown).  Moreover the longitudinal resistance ratio has risen
to R$_{[011]}$/R$_{[01\overline{1}]}$ $\sim$ 16.  

Figure 3 summarizes the rapid low-temperature evolution of the
resistance at $\nu$=7/2 from T=275mK down to T=50mK.  Above
T$\sim$150mK the resistance along both directions is approximately
125$\Omega$ and shows litte variation with temperature.  Below
T=150mK the resistances change dramatically.  Between T=150mK and
50mK the resistance along [011] increases by a factor of 7 and
shows no indication of saturation at T=50mK.  Conversely, the
resistance along [01$\overline{1}$] has fallen by a factor of 2
from its value at T=150mK and also does not appear saturated.

\begin{figure}
\includegraphics[width=.9\columnwidth]{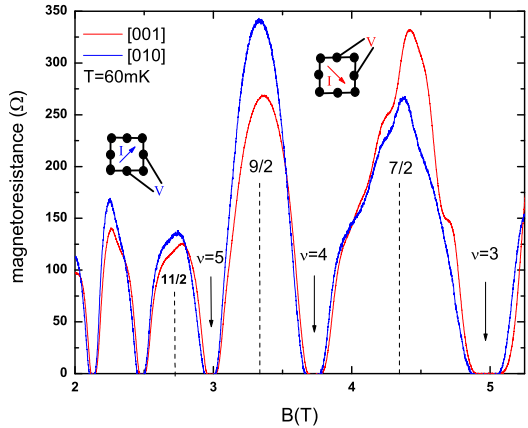}
 \caption{Resistance
at T=60mK along the [001] (red) and the [010] (blue) direction.
The transport is isotropic at all filling factors between $\nu$=3
and $\nu$=6.}
\end{figure}

The observation of isotropic transport for 2D holes at $\nu$=9/2
flanked by anisotropic transport at $\nu$=11/2 and $\nu$=7/2 is
the most striking feature of this study.  In 2D electron systems
which display anisotropic transport, the anisotropy only resides
in the N$\geq$2 LL's and also shows the largest resistance ratio
at $\nu$=9/2.  In addition, no alternating sequence of
anisotropic/isotropic states is observed.  Given these unexpected
result for 2D holes one can ask if perhaps $\nu$=9/2 is still
anisotropic for holes, albeit with  different principle axes than
the [011] and [01$\overline{1}$] directions that prevail at
$\nu$=7/2 and $\nu$=11/2.  In order to investigate this
possibility, we measured the resistance with current flowing
primarily along the [001] and [010] directions.  This
configuration amounts to flowing current between two corner
contacts, along the diagonal of the square sample.  The voltage is
monitored using two contacts at the midpoint of sample edges (see
Fig. 4).  The transport at $\nu$=9/2 remains isotropic indicating
that the principle axes of anisotropy has not switched in moving
from $\nu$=7/2 to $\nu$=9/2.  In addition we observe that the
initial anisotropies observed at $\nu$=11/2 and 7/2 are lost with
current flow along the [010] and [001] directions, as expected.

Although 2D electron systems do not exhibit anisotropic transport in the
N=1 Landau level at $\nu$=5/2 and $\nu$=7/2 in a perpendicular
magnetic field, Lilly \cite{mike2} and Pan
\cite{Pan2} have observed that the incompressible quantum Hall
states at $\nu$=7/2 and $\nu$=5/2 are replaced by compressible
anisotropic states under the application of large tilting angles
corresponding to in-plane magnetic field $\sim$ 8T.
These results  suggest that the physics influencing
the formation of compressible striped phases in the N$\geq$2 LL's
may be active in the N=1 LL under the appropriate change of the
effective interaction induced by the large in-plane field. In
numerical studies, Rezayi and Haldane (RH) \cite{rezayi1} have
shown that the incompressible quantum Hall state at $\nu$=5/2 is
near to a phase transition into a compressible striped phase. In
the psuedopotential formulation of the FQHE \cite{Haldane1}, the
nature of the ground state is found to depend sensitively on the
relative strengths of the pseudopototential parameters $V_1$ and
$V_3$, where $V_m$ is the energy of a pair of electrons in a state
of relative angular momentum $m$.  RH find that at $\nu$=5/2,
small variations in $V_1$ and $V_3$ can drive the phase transition
and suggest that the proximity of the critical point to the
Coulomb potential is the principle reason that transport becomes
anisotropic in the tilting experiments.
\begin{figure}
\includegraphics[width=.9\columnwidth]{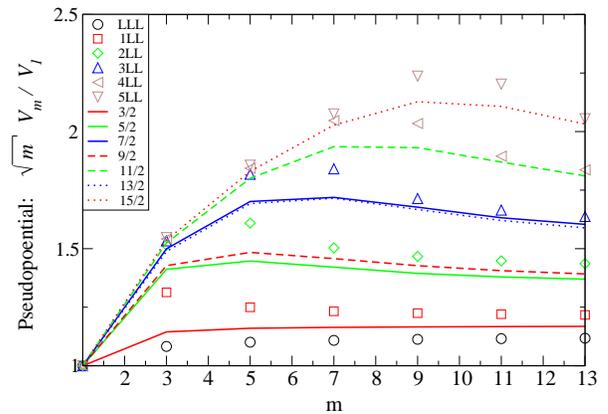}
 \caption{Ratio of psuedopotentials V$_m$/V$_1$ multiplied by $\sqrt{m}$ for the
first 5 electron Landau levels (open symbols) assuming an
infinitely thin quantum well. Also shown is the same ratio
($\sqrt{m}$V$_m$/V$_1$) for the self-consistently calculated  LL
structure for our 2DHS at several values of half-filling (solid
lines).}
\end{figure}

What distinguishes our 2D hole system from the 2D electron system
such that the fractional quantum Hall state at $\nu$=7/2 is
destabilized and replaced by an compressible anisotropic state and
the transport at $\nu$=9/2 remains isotropic rather than
displaying the anisotropy seen in electron systems?  We claim that
the strong spin-orbit coupling in the 2DHS is the critical
difference.   Spin-orbit coupling strongly mixes valence band
states, which alters the orbital structure of hole Landau levels at
B$\neq$0 \cite{Yang1, Ekenberg1}.  The nature of the single
particle wavefunctions that comprise a given LL alters the
pseudopotential parameters, significantly influencing the
correlations among the holes \cite{MacDonald1,Yang2}.  Following
Ref. \cite{Yang2}, we have self-consistently calculated the Landau
level structure in the Hartree approximation (while keeping axial
terms as in \cite{Yang1}). Within this basis of single-particle
states, the pseudopotential parameters for our 2DHS are then
calculated. The 4x4 Luttinger Hamiltonian is used to describe the
four highest valence bands and the effects of band anisotropy are
included in the Luttinger parameters \cite{roland1}.  All of the
valence LL's are found to be strongly mixed.  At $\nu$=7/2, the
valence LL is comprised largely of the N=2 (50\%) and N=4 (35\%)
oscillator functions.  At $\nu$=9/2 the dominant oscillator
components are N=1 (50\%), N=2 (13\%), and N=3 (24\%).  At
$\nu$=11/2, N=3 (49\%) and N=5 (36\%) are most heavily weighted.
The results of the pseudopotential calculations are shown in Fig.
5, plotted as $\sqrt{m}V_m/V_1$. For comparison, we also plot
$\sqrt{m}V_m/V_1$ for the first five {\it electron} Landau levels
(assuming an infinitely thin quantum well). Interestingly, the
pseudopotential structure for our 2DHS differs significantly from
the calcuation for electron LL's at several key filling factors.
We begin consideration at $\nu$=11/2. For our 2DHS the
pseudopotential structure at small $m$ is quite similar to the N=3
and N=4 electron LL's.  The anisotropic transport seen at
$\nu$=11/2 in our 2DHS is thus consistent with the known
correlations in the higher electron LL's \cite{Ru1,mike1}.  At
$\nu$=9/2 a quite remarkable change is seen. The pseudopotential
ratios $\sqrt{3}V_3/V_1$ and $\sqrt{5}V_5/V_1$ are appreciably
below the N=2 electron LL, possessing ratios between N=1 and N=2
values. Our calculation indicates that 2DHS LL at $\nu$=9/2 has
acquired correlations, at least in part, associated with N=1 LL,
consistent with our observation of isotropic transport at
$\nu$=9/2.  At $\nu$=7/2, an even more dramatic difference between
the electron and hole correlations is manifest.  All of the
pseudopotential ratios for $\nu$=7/2 are far above those of the
N=1 electron LL. The calculated pseudopotential ratios rest
between the N=2 and N=3 electron LL's and are larger than those
calculated at $\nu$=9/2. The calculated pseudopotential ratios at
$\nu$=7/2 suggest an origin to the highly anisotropic transport
observed in our 2DHS at $\nu$=7/2. The correlations operational at
$\nu$=7/2 are in fact more closely related to the physics of the
N$\geq$2 Landau levels than the N=1 level.  Thus our calculation
of the pseudopotential ratios specific to our 2DHS provide, at
least qualitatively, a consistent picture of the alternating
sequence of anisotropic/isotropic transport observed in
experiment.  We note that a similar analysis might elucidate the transport anomolies investigated in Ref.\cite{mansour1}.

In conclusion, we observe an alternating sequence of
anisotropic/isotropic transport at filling factors $\nu$=11/2,
9/2, and 7/2 in a high quality 2DHS.  These results are quite
distinct from the known behavior of 2D electron systems.
The transport experiments combined with calculations of the Landau level structure and pseudopotential
parameters of our 2DHS indicate that the type of correlated ground state observed at a
particular filling factor depends sensitively on the nature of the
single particle states available to the system.  The great
flexibility to tune the Landau level structure of a 2DHS through
sample design suggests future experiments in which the correlated
ground state observed at a given filling factor is constructed by
judicious choice of 2DHS sample parameters.

M. J. Manfra thanks R. L. Willett for helpful conversations.  Z.  Jiang is supported by NSF
under DMR-03-52738 and by the DOE under DE-AIO2-04ER46133.

\end{document}